\documentclass[epj]{svjour}
\usepackage{graphics}
\usepackage{mathrsfs}  
\usepackage{adjustbox}
\usepackage[all]{xy}
\usepackage{epsfig, cancel}
\usepackage{cite}
\usepackage{ulem}
\usepackage{latexsym}
\usepackage{url}
\usepackage{dcolumn}
\usepackage{color}
\usepackage{amsfonts,amssymb,amsmath, txfonts}
\usepackage{graphicx,epsfig}
\usepackage{psfrag}
\usepackage{subfigure}
\usepackage{hyperref}
\hypersetup{colorlinks=true}
\usepackage{mathtools}
\usepackage{enumitem}
\usepackage{float}
\usepackage[dvipsnames]{xcolor}
\usepackage{xcolor}
\hypersetup{ linktoc=all,
    colorlinks, linkcolor={brightpink},
    citecolor={blue}, urlcolor={blue}
}
\definecolor{rosy}{RGB}{230,235,252}
\definecolor{myframetitle}{RGB}{90,89,170}
\definecolor{myblocktitle}{RGB}{140,185,249}
\definecolor{mytitle}{RGB}{10,80,26}

\definecolor{darkgreen}{RGB}{27,130,45}
\definecolor{darkblue}{rgb}{0,0,0.3}
\definecolor{darkred}{rgb}{0.7,0,0}

\definecolor{light gray}{RGB}{220,220,220}
\definecolor{dark purple}{RGB}{108,0,217}
\definecolor{pink}{RGB}{190,20,100}
\definecolor{orang}{RGB}{193,63,0}
\definecolor{green}{RGB}{11,98,17}
\definecolor{darkpink}{RGB}{153,0,76}
\definecolor{bluegreen}{RGB}{0,102,102}
\definecolor{greenlagan}{RGB}{0,102,0}
\definecolor{redgreen}{RGB}{102,102,0}
\definecolor{Redgreen}{RGB}{153,76,0}
\definecolor{vividviolet}{rgb}{0.62, 0.0, 1.0}
\definecolor{amaranth}{rgb}{0.9, 0.17, 0.31}
\definecolor{palatinateblue}{rgb}{0.15, 0.23, 0.89}
\definecolor{brightpink}{rgb}{1.0, 0.0, 0.5}
\definecolor{cornflowerblue}{rgb}{0.39, 0.58, 0.93}
\definecolor{deepcarminepink}{rgb}{0.94, 0.19, 0.22}
\definecolor{radicalred}{rgb}{1.0, 0.21, 0.37}
\def\H0{{\text{H}\hspace*{-2.05mm}\text{H} 0\hspace*{-1.35mm}0\ }}

\def\be{\begin{equation}}
\def\ee{\end{equation}}
\def\beq{\begin{equation}}
\def\eeq{\end{equation}}
\def\bea{\begin{eqnarray}}
\def\eea{\end{eqnarray}}

\newcommand{\nn}{\nonumber \\}

\begin{document}
\title{On Redshift Evolution \& Negative Dark Energy Density in Pantheon+ Supernovae}

\author{M. Malekjani\inst{1} \and R. Mc Conville\inst{2} \and E. \'O Colg\'ain\inst{2} \and S. Pourojaghi \inst{3} \and M. M. Sheikh-Jabbari \inst{3}% etc
}                     % Do not remove

\institute{Department of Physics, Bu-Ali Sina University, Hamedan 65178, Iran \and Atlantic Technological University, Ash Lane, Sligo, Ireland \and 
School of Physics, Institute for Research in Fundamental Sciences (IPM), P.O.Box 19395-5531, Tehran, Iran}
\date{Received: date / Revised version: date}
\abstract{
Within  the Friedmann-Lema\^itre-Robertson-Walker (FLRW) framework, the Hubble constant $H_0$ is an integration constant. Thus, consistency {of the model demands observational constancy of $H_0$.} We demonstrate redshift evolution of best fit $\Lambda$CDM parameters $(H_0, \Omega_{m})$ in Pantheon+ supernove ({SNe}). Redshift evolution of best fit cosmological parameters is a prerequisite to finding {a statistically significant evolution  as well as identifying alternative models that are competitive with $\Lambda$CDM in a Bayesian model comparison}. To assess statistical significance, we employ three different methods: i) Bayesian model comparison, ii) mock simulations and iii) profile distributions. {The first shows a marginal preference for the vanilla $\Lambda$CDM model over an ad hoc model with 3 additional parameters and an unphysical jump in cosmological parameters at $z=1$. From mock simulations}, we estimate the statistical significance of redshift evolution of best fit parameters and negative dark energy density ($\Omega_m > 1$) to be in the $1-2 \sigma$ range, depending on the criteria employed. Importantly, in direct comparison to the same analysis with the earlier Pantheon sample we find that statistical significance {of redshift evolution of best fit parameters} has increased, as expected for a physical effect. Our profile distribution analysis {demonstrates a shift in $(H_0, \Omega_m)$ in excess of $95\%$ confidence level for SNe with redshifts $z > 1$ and also shows} that a degeneracy in MCMC posteriors is not equivalent to a curve of constant $\chi^2$. Our findings can be interpreted as a statistical fluctuation or unexplored systematics in Pantheon+ or $\Lambda$CDM model breakdown. The first two possibilities are disfavoured by similar trends in independent probes.
\PACS{
      {98.80.Es}{Observational cosmology }   \and
      { 95.35.+d}{Dark energy}
     } % end of PACS codes
} %end of abstract
\maketitle

\section{Introduction}
\label{sec:intro}
The (flat) $\Lambda$CDM cosmological model is an extremely successful minimal model that returns seemingly consistent cosmological parameters across Type Ia supernovae ({SNe}) \cite{Riess:1998cb, Perlmutter:1998np}, Cosmic Microwave Background \cite{Planck:2018vyg} and baryon acoustic oscillations \cite{Eisenstein:2005su}. Despite this success, comparison of early and late Universe cosmological parameters has revealed discrepancies \cite{Riess:2021jrx, Freedman:2021ahq, Pesce:2020xfe, Blakeslee:2021rqi, Kourkchi:2020iyz, DES:2021wwk, KiDS:2020suj}. The origin \cite{Perivolaropoulos:2021jda, Abdalla:2022yfr} and resolution \cite{DiValentino:2021izs,Schoneberg:2021qvd} of these anomalies is a topic of debate. We observe that the $\Lambda$CDM model describes approximately 13 billion years of evolution {of the Hubble parameter $H(z)$} in the late Universe (conservatively redshifts $z \lesssim 30$) with a single fitting parameter, matter density today $\Omega_m$.\footnote{The Hubble constant $H_0$ is merely the scale of $H(z)$, so it does not dictate evolution. $\Omega_m$ is also in effect an integration constant from the continuity equation.} Objectively, given the prevailing belief that $\Omega_m \sim 0.3$, this marks billions of years of %\sout{background cosmology} 
evolution with effectively no free parameters. 
%If true, this is miraculous.  

As originally pointed out \cite{Krishnan:2020vaf} (see also \cite{Krishnan:2022fzz}), within the FLRW framework, any mismatch between $H(z)$, an unknown function inferred from Nature, and a theoretical assumption on the effective EoS $w_{\textrm{eff}}(z)$, e. g. the $\Lambda$CDM model, must mathematically lead to a Hubble constant $H_0$ that evolves with effective redshift. 
%Unless one has a precise model within the FLRW setting, it 
Simply put, a redshift-dependent $H_0$ is indicative of a bad model  \cite{Krishnan:2020vaf}. This prediction can be tested in the late Universe, where the $\Lambda$CDM model reduces to two fitting parameters: 
\be
\label{lcdm}
H(z) = H_0 \sqrt{1-\Omega_m + \Omega_m (1+z)^3}.
\ee

To date, independent studies have documented decreasing $H_0$ trends within model (\ref{lcdm}) across strong lensing time delay (SLTD) \cite{Wong:2019kwg, Millon:2019slk}, Type Ia supernovae ({SNe}) \cite{Dainotti:2021pqg, Dainotti:2022bzg, Colgain:2022nlb, Colgain:2022rxy, Dainotti:2023yrk,Yu:2022wvg} and combinations of cosmological data sets \cite{Krishnan:2020obg, Jia:2022ycc, Hu:2022kes}. Moreover, quasar (QSO) Hubble diagrams \cite{Risaliti:2015zla, Risaliti:2018reu, Lusso:2020pdb} show a preference for larger than expected $\Omega_m$ values, $\Omega_m \gtrsim 1$ \cite{Yang:2019vgk, Khadka:2020vlh,Khadka:2020tlm, Khadka:2021xcc}. It was subsequently noted that $\Omega_m$ increases with effective redshift in {SNe} and QSO samples \cite{Colgain:2022nlb, Pourojaghi:2022zrh, Pasten:2023rpc}. Although the trend in any given observable is not overly significant, e. g. $\lesssim 2 \sigma$ for SLTD \cite{Wong:2019kwg, Millon:2019slk}, combining probabilities from independent observables using Fisher's method, the significance increases quickly \cite{Colgain:2022rxy}.  

Simple binned mock $\Lambda$CDM data analysis \cite{Colgain:2022rxy, Colgain:2022tql} suggests that evolution of $(H_0, \Omega_m)$ best fit parameters must be expected in any data set that only provides either observational Hubble $H(z)$ or angular diameter $ D_{A}(z)$ or luminosity distance $D_{L}(z)$ constraints. \footnote{Note that $D_L(z)$ and $D_{A}(z)$ are not independent within FLRW setting, since $D_{L}(z) = (1+z)^2 D_{A}(z)$.} If true, one can expect to separate any given sample into low and high redshift subsamples and see discrepancies in the $(H_0, \Omega_m)$-plane. Here, we highlight the feature in the latest Pantheon+ {SNe} sample \cite{Brout:2022vxf, Scolnic:2021amr}.\footnote{We note that Pantheon+ is statistically poorer than the Pantheon sample \cite{Pan-STARRS1:2017jku} at higher redshifts; 57 higher redshift SNLS \cite{SNLS:2011lii, SNLS:2011cra} {SNe} beyond $z=0.8$ have been removed due to potential evolution in inferred distances \cite{Brout:2021mpj}.} The main message of this letter is that evolution of $(H_0, \Omega_m)$ with effective redshift persists in Pantheon+ {SNe}. Furthermore, an increasing $\Omega_m$ trend, evident at higher redshifts, continues beyond $\Omega_m=1$  giving rise to negative DE densities at $z \gtrsim 1$. In light of concerns highlighted in \cite{Davis:2019wet, Rameez:2019wdt,Steinhardt:2020kul}, the Pantheon+ sample improves on redshift corrections \cite{Carr:2021lcj}. Thus, errors in the handling of redshifts can be precluded as the origin of the trend. It is worth stressing again that \cite{Colgain:2022rxy, Colgain:2022tql} provide a mathematical proof that redshift evolution of best fit $\Lambda$CDM parameters cannot be ruled out in mock Planck-$\Lambda$CDM data. There are then two relevant questions. Is redshift evolution of best fit $\Lambda$CDM parameters evident in observed data? If so, what is its statistical significance?

{In cosmology the default is to assess statistical significance with Markov Chain Monte Carlo (MCMC)}. The increasing $\Omega_m$ trend is evident in MCMC posteriors, but as we demonstrate, the $H_0$ posterior is subject to projection effects due to a degeneracy (banana-shaped contour) in the 2D $(H_0, \Omega_m)$ posterior. In the literature, this is interpreted as the data failing to constrain the model, but as we will show in section \ref{sec:PD}, this is a misconception because it is not supported by the $\chi^2$ {(see also \cite{Colgain:2023bge})}. We overcome the MCMC degeneracy in three complementary ways. First, we provide a Bayesian comparison between the $\Lambda$CDM model and the $\Lambda$CDM model with a split at redshift $z_{\textrm{split}}$, where $(H_0, \Omega_m)$ are allowed to adopt different values at low and high redshift. Secondly, we employ a frequentist comparison between best fits of the observed data and mock data that focuses on different criteria quantifying evolution in the sample. Finally, we analyse the $\chi^2$ through profile distributions. For the Pantheon+ sample split at $z_{\textrm{split}}=1$ {we find a shift in the cosmological parameters that exceeds 95\% confidence level}. Note, Pantheon+ is presented as a sample in the redshift range $0 < z \leq 2.26$, but redshift evolution of cosmological parameters is evident in the $\Lambda$CDM model from $z = 0.7$ onwards. 

Hints of negative DE densities, especially at higher redshifts, are widespread in the literature, so our observations in Pantheon+ may be unsurprising. Indeed, while $\Lambda$CDM mock analysis in \cite{Colgain:2022rxy, Colgain:2022tql} confirms that $\Omega_m > 1$ best fits are precluded with low redshift data, this is no longer true at higher redshifts. We stress again that this is a purely mathematical feature of the $\Lambda$CDM model. Starting with studies incorporating Lyman-$\alpha$ BAO \cite{Aubourg:2014yra}, one of the first observables discrepant with Planck-$\Lambda$CDM \cite{BOSS:2014hwf, BOSS:2013igd, duMasdesBourboux:2020pck}, claims of negative DE densities at higher redshifts, including anti-de Sitter (AdS) vacua at high redshift \cite{Dutta:2018vmq, Sen:2021wld} (however see \cite{Visinelli:2019qqu}) \footnote{AdS vacua pre-recombination resolve Hubble tension without invoking $H_0$ priors  
\cite{Ye:2020btb, Ye:2020oix, Wang:2022jpo, Wang:2022nap, Jiang:2022uyg}.} and features in data reconstructions \cite{Mortsell:2018mfj, Poulin:2018zxs, Wang:2018fng, Bonilla:2020wbn, Escamilla:2021uoj}, have been noticeable.\footnote{One cannot get negative DE densities when one works with a barotropic EoS $w_{\textrm{DE}}(z)$, but one may find null DE densities by sending $w_{\textrm{DE}}(z) \rightarrow - \infty$ \cite{Sahni:2014ooa, Ozulker:2022slu}. More generally, one notes a preference for $w_{\textrm{DE}}(z) < 0$ at higher $z$ \cite{Zhao:2017cud, Capozziello:2018jya}.} This has led to extensive attempts to model negative DE densities \cite{DiValentino:2017rcr, Banihashemi:2018oxo, Banihashemi:2018has, Akarsu:2019ygx, Perez:2020cwa, Akarsu:2020yqa, Calderon:2020hoc, Acquaviva:2021jov, LinaresCedeno:2021aqk, Akarsu:2022lhx, Moshafi:2022mva}, most simply as sign switching $\Lambda$ models \cite{Akarsu:2019hmw, Akarsu:2021fol, Akarsu:2022typ, DiGennaro:2022ykp, Ong:2022wrs}. Given the sparseness of {SNe} data beyond $z=1$, claims of negative DE densities are usually attributed to Lyman-$\alpha$ BAO,\footnote{{$\Omega_m > 1$ best fits also appear in high redshift observational Hubble data (OHD) \cite{Colgain:2022rxy} that incorporates historical Lyman-$\alpha$ BAO \cite{Aubourg:2014yra}. However, when updated to the latest Lyman-$\alpha$ BAO constraints \cite{Neveux:2020voa, duMasdesBourboux:2020pck}, one finds a $\Omega_m < 1$ best fit, admittedly one that still precludes the Planck $\Omega_m$ value at greater than $95 \%$ confidence level \cite{Colgain:2023bge}. Thus, improvements in data quality can remove signatures of negative DE densities by bringing $\Omega_m$ values back closer to Planck values.}} but here we see the same feature in state of the art Pantheon+ {SNe}. It is plausible that selection effects are at play (see discussion in \cite{Dainotti:2021pqg}), but if the arguments in \cite{Colgain:2022rxy, Colgain:2022tql} hold up, then $\Omega_m > 1$ $\Lambda$CDM best fits to data in high redshift bins cannot be precluded. On the contrary, they can be expected. 

\section{Preliminaries}
\label{sec:warmup}
\begin{figure}[htb]
\centering
\includegraphics[width=80mm]{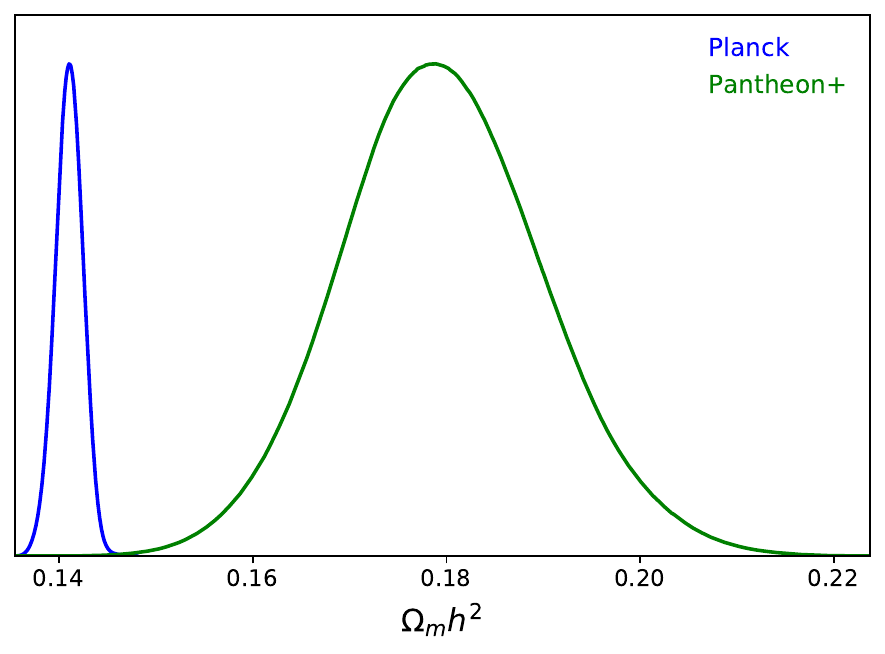} 
\caption{{$3.7 \sigma$ tension between Planck and Pantheon+ for the parameter combination that dictates the high redshift behaviour of the Hubble parameter $H(z)$ in the late Universe. We made use of \textit{GetDist} \cite{Lewis:2019xzd}.}}
\label{fig:tension} 
\end{figure}

Our analysis starts by following and recovering results in \cite{Perivolaropoulos:2023iqj} (see also \cite{Brout:2022vxf}). We set the stage with a preliminary consistency check. In short, we extremise the likelihood, 
\be
\label{chi2}
\chi^2 = \vec{Q}^{T} \cdot (C_{\textrm{stat+sys}})^{-1} \cdot \vec{Q}, 
\ee
where $\vec{Q}$ is a 1701-dimensional vector and $C_{\textrm{stat+sys}}$ is the covariance matrix of the Pantheon+ sample \cite{Brout:2022vxf}. The Pantheon+ sample has 1701 SN light curves, 77 of which correspond to galaxies hosting Cepheids in the low redshift range $0.00122 \leq z \leq 0.01682$. In order to break the degeneracy between $H_0$ and the absolute magnitude $M$ of Type Ia, we define the vector 
\be
Q_i = 
\begin{cases} m_i -M - \mu_i,  \quad i \in \textrm{Cepheid hosts}  \\
m_i - M - \mu_{\textrm{model}}(z_i), \quad \textrm{otherwise}
\end{cases}
\ee
where $m_i$ and $\mu_i \equiv m_i - M$ denote the apparent magnitude and distance modulus of the $i^{\textrm{th}}$ SN, respectively. The cosmological model, which we assume to be the $\Lambda$CDM model (\ref{lcdm}), enters through the following relations: 
\bea
\mu_{\textrm{model}}(z) &=& 5 \log \frac{d_{L}(z)}{\textrm{Mpc}} + 25, \nn
d_{L}(z) &=& c (1+z) \int_{0}^{z} \frac{\textrm{d} z^{\prime}}{H(z^{\prime})} .  
\eea
Extremising the likelihood, one arrives at the best fit values, 
\be
H_0 = 73.42 \textrm{ km/s/Mpc}, \quad
\Omega_m = 0.333, \quad M = -19.248, 
\ee
which are in perfect agreement with \cite{Perivolaropoulos:2023iqj}. {We estimate the $1 \sigma$ confidence intervals through an MCMC exploration of the likelihood with \textit{emcee} \cite{Foreman-Mackey:2012any}, finding excellent agreement with Fisher matrix analysis \cite{Perivolaropoulos:2023iqj},}  
\be\label{full_sample_constraints}
\begin{split}
H_0 &=  73.41^{+0.97}_{-1.00} \textrm{ km/s/Mpc}, \\
\Omega_m &= 0.333^{+0.018}_{-0.017}, \quad M = - 19.248^{+0.028}_{-0.030}. 
\end{split}
\ee
It is interesting to compare Pantheon+ constraints on $\Omega_m h^2$ $(h:= H_0/100)$ directly with Planck. In Fig. \ref{fig:tension} we highlight a $3.7 \sigma$ tension,\footnote{The relevant constraints are $\Omega_m h^2 = 0.1430 \pm 0.0014$ and $\Omega_m h^2 = 0.18 \pm 0.01$ for Planck and Pantheon+, respectively.} which importantly impacts the high redshift behaviour of $H(z) \sim H_0 \sqrt{\Omega_m} (1+z)^{\frac{3}{2}}$ in the late Universe. This is interesting, as we start to see evolution in best fit $\Lambda$CDM parameters at higher redshifts. {Given the tension in the Hubble constant \cite{Riess:2021jrx, Freedman:2021ahq, Pesce:2020xfe, Blakeslee:2021rqi, Kourkchi:2020iyz}, our focus here is on $H_0$ and by extension $\Omega_m$, since both parameters are correlated when one fits data. Of course, if the fitting parameters $H_0$ and $\Omega_m$ change with effective redshift, there is no guarantee that $\Omega_m h^2$ is a constant. If the constancy of $\Omega_m h^2$ can be tested, this allows one to study the asumption that matter is pressureless. Such studies will require data exclusively in the matter dominated regime where DE and radiation sectors are irrelevant. Given the sparcity of high redshift $z > 1$ data, competitive studies are still a few years off.}

\section{Splitting Pantheon+}
Having confirmed the results quoted in \cite{Perivolaropoulos:2023iqj}, we depart from earlier analysis and crop the Pantheon+ covariance matrix in order to isolate the $77 \times 77$-dimensional covariance matrix $C_{\textrm{Cepheid}}$ corresponding to {SNe} in Cepheid host galaxies and define a new likelihood that is only sensitive to the absolute magnitude $M$, 
\bea
\chi_{\textrm{Cepheid}}^2 &=& (\vec{Q}^{\prime})^{T} \cdot (C_{\textrm{Cepheid}})^{-1} \cdot \vec{Q}^{\prime}, \nn
Q_i^{\prime} &=& m_i - M - \mu_i, \quad i \in \textrm{Cepheid hosts}.  
\eea
We can now split the remaining 1624 {SNe}  into low and high redshift samples, which we demarcate through a redshift $z_{\textrm{split}}$. One can crop the original covariance matrix accordingly to get $C_{\textrm{SN}}$ for either the low or high redshift sample, but we will primarily focus on the high redshift subsample with $z > z_{\textrm{split}}$. The reason being that {SNe} samples have a low effective redshift, $z_{\textrm{eff}} \sim 0.3$, and it is well documented that Planck values $\Omega_m \sim 0.3$ are preferred. The hypothesis we explore is that such results overlook evolution at higher redshifts, so this explains the focus on high redshift subsamples. In summary, we study the new likelihood, 
\be
\label{chi2_split}
\chi^2 = \chi^2_{\textrm{Cepheid}} + \chi_{\textrm{SN}}^2, 
\ee
where we have defined, 
\be
\label{chi2_split_details}
\begin{split}
\chi_{\textrm{SN}}^2 &= (\vec{\tilde{Q}})^{T} \cdot (C_{\textrm{SN}})^{-1} \cdot \vec{\tilde{Q}}, \\
\tilde{Q}_i &= m_i - M - \mu_{\textrm{model}}(z_i). 
\end{split}\ee
The redshift range of the Pantheon+ sample \cite{Brout:2022vxf} is $0.00122 \leq z \leq 2.26137$, so we take $z_{\textrm{split}}$ in this range. {In the next section we begin the tomographic analysis of splitting the Pantheon+ sample into a low and high redshift subsample.} {We remark that the likelihoods presented in equation (\ref{chi2}) and (\ref{chi2_split}) omit a constant normalisation. Being a constant, it plays no role when one fits data, and is thus routinely omitted in the literature \cite{Brout:2022vxf}. However, this term is relevant when one performs Bayesian model comparison. We will reinstate the normalisation later.}

\section{Analysis}
It is widely recognised that confronting exclusively high redshift SNe data to the $\Lambda$CDM model, MCMC inferences are typically impacted by degeneracies, i. e.  banana-shaped posteriors, in the $(H_0, \Omega_m)$-plane. Later we confirm the impact of projection effects on MCMC posteriors as priors are relaxed in the presence of a degeneracy.\footnote{{In the cosmology literature, it is routinely assumed that all points within MCMC $68 \%$ credible intervals provide an equally good fit to the data, even in the presence of a degeneracy. As demonstrated in \cite{Colgain:2023bge}, this need not be the case; one can find settings where frequentist confidence intervals are constrained, but MCMC credible intervals are at best inconclusive, thereby undermining any analysis that rests only on MCMC.}} We overcome the degeneracy in MCMC marginalisation through three different prongs of attack {that only rest upon on the likelihood or $\chi^2$. Here it is worth noting that MCMC is merely an algorithm, whereas the $\chi^2$ is a measure or metric of how well a point in parameter space fits the data.} First, we provide a Bayesian model comparison based on the Akaike Information Criterion (AIC) \cite{AIC} between the $\Lambda$CDM model and a $\Lambda$CDM model allowing a jump in cosmological parameters $(H_0, \Omega_m)$ at a fixed redshift. {Despite the vanilla $\Lambda$CDM model being preferred by the AIC, the analysis demonstrates that an alternative model, even a physically ad hoc model that contradicts the basic fundamentals of FLRW, becomes more competitive if the $\Lambda$CDM fitting parameters change with effective redshift when confronted to data.} Secondly, in a frequentist analysis we resort to a comparison between best fits of observed and mock data in the same redshift range with the same data quality to ascertain the significance of evolution. {Finally, later in section \ref{sec:PD}, we employ profile distributions as a secondary frequentist approach.}
%that by construction track the minimum of the $\chi^2$.

\subsection{Bayesian Interpretation}
\label{sec:bayesian}
{One may interpret the results in Table \ref{tab:zsplit_H0_Om} as a comparison between two models. The first is the $\Lambda$CDM model fitted over the entire redshift range of the {SNe}, $0.00122 \leq z \leq 2.26137$, with three parameters $(H_0, \Omega_m, M)$, while the second is the $\Lambda$CDM model with a split at redshift $z_{\textrm{split}}$ allowing the model to adopt different values of $(H_0, \Omega_m)$ above and below the split. Note that the likelihood (\ref{chi2_split}) separates {SNe} in Cepheid host galaxies and their only role is to constrain $M$. For this reason, one is only fitting two effective parameters $(H_0, \Omega_m)$. Furthermore, by introducing the data split, we are comparing this effective two parameter model $(H_0, \Omega_m)$ with an effective five parameter model $(H^{(1)}_0, \Omega^{(1)}_m, H^{(2)}_0, \Omega^{(2)}_m, z_{\textrm{split}})$. {Table \ref{tab:zsplit_H0_Om} presents improvements in the $\chi^2$ without the normalisation corresponding to the logarithm of the determinant of the covariance matrix $C_{\textrm{stat+sys}}$. Since we truncate out $C_{\textrm{stat+sys}}$ entries when we split the SNe, this increases the normalisation, thereby penalising the model with the split beyond the 3 extra parameters introduced. We will quantify this number in turn, but only in competitive settings relative to the $\Lambda$CDM model where the improvement in $\chi^2$ in Table \ref{tab:zsplit_H0_Om} is enough to overcome the additional parameters, i. e. $\Delta \chi^2 < -6$. \footnote{{Whenever the covariance matrix $C$ has large dimensionality and small entries, it is difficult to determine its determinant $|C|$; beyond a certain dimensionality, one encounters $|C| = 0 \Rightarrow  \ln |C| = -\infty$ within machine precision. Nevertheless, when the dimensionality becomes smaller, a finite, non-zero determinant is calculable.}}}

{It should be noted that while model $A$ is the vanilla $\Lambda$CDM model, the model $B$ that serves as a foil to $\Lambda$CDM is a contradiction, because if $H_0$ and $\Omega_m$ change with effective redshift, this violates the mathematical requirement that both are integration constants. For this reason, model $B$ could never replace $\Lambda$CDM. Nevertheless, the result is instructive as Bayesian model comparison is prevalent in the cosmology literature. That being said, the focus of this paper is performing a consistency check of the $\Lambda$CDM model and this does not necessitate a model $B$. What the analysis here shows is that we are getting close to a point in time where models incorporating evolution in the fitting parameters $H_0$ and $\Omega_m$ may be more competitive than $\Lambda$CDM, simply based on SNe data alone.}

We recall the Akaike Information Criterion (AIC) \cite{AIC}, 
\be
\textrm{AIC} = {-2 \ln \mathcal{L}_{\textrm{max}} +2 d = \ln |C_{\textrm{stat+sys}}| +\chi_{\textrm{min}}^2+2 d}, 
\ee
where $\chi_{\textrm{min}}^2$ is the minimum of the $\chi^2$, $d$ is the number of free parameters {and $|C_{\textrm{stay+sys}}|$ denotes the determinant of the Pantheon+ covariance matrix $C_{\textrm{stat+sys}}$. Since the latter is a constant, it has no bearing on the best fit parameters, but it impacts the AIC analysis. \footnote{{We thank an anonymous EPJC referee for pointing this important point out.}} However, since $C_{\textrm{stat+sys}}$ is a large matrix with small numerical entries, determining the absolute value of $|C_{\textrm{stat+sys}}|$ within machine precision is difficult. One can simplify the problem by noting that the $77 \times 77$ matrix $C_{\textrm{Cepheid}}$ is common to both the $\Lambda$CDM model and the $\Lambda$CDM model with a jump in cosmological parameters, so it contributes to both AIC values and drops out. Thus, we only need to the study the $1624 \times 1624$ covariance matrix $C_{\textrm{SN}}$, but this is still a large matrix with small numerical entries.} 

{Since the $\Lambda$CDM model with a jump in cosmological parameters necessitates three additional parameters, i. e. $\Delta d = 3$, this penalty can only be absorbed to give a lower AIC if $\Delta \chi_{\textrm{min}}^2 < -6$. The results of splitting the Pantheon+ sample and fitting the $\Lambda$CDM model to data below and above $z = z_{\textrm{split}}$ are shown in Table \ref{tab:zsplit_H0_Om}. We find that refitting the low redshift sample typically leads to small improvements in $\chi^2$, whereas refits of the high redshift sample lead to greater improvements. This outcome is expected if there is evolution across the sample; the evolution is only expected at higher redshifts because SNe samples have a low effective redshift, and as we have noted, SNe samples generically prefer Planck values $\Omega_m \sim 0.3$. In particular, $z_{\textrm{split}}=1$ gives rise to greatest reduction in $\chi^2_{\textrm{min}}$ with respect to the $\Lambda$CDM model without the split. However, we need to make sure that differences in $\ln |C_{\textrm{SN}}|$ do not counter the improvement in $\chi_{\textrm{min}}^2$.}

{To that end, consider 
\be
C_{\textrm{SN}} = \left( \begin{array}{cc} A & B \\ B^{T} & C \end{array} \right)
\ee
where $A, B$ and $C$ are respectively $1599 \times 1599, 1599 \times 25$ and $25 \times 25$-dimensional matrices. Note that the dimensionalities are fixed by the choice of $z_{\textrm{split}}=1$. The determinant of this block diagonal matrix is 
\be
|C_{\textrm{SN}}| = |A| \cdot |C-B^{T} A^{-1} B|,  
\ee
provided the matrix $A$ is invertible.
Note that when one introduces the split at $z_{\textrm{split}}=1$, one sets $B = 0$. As a result, the difference in the $\ln |C_{\textrm{SN}}|$ is 
\bea
\Delta  \ln |C_{\textrm{stat+sys}}| &=& \Delta \ln |C_{\textrm{SN}}|, \nn &=& \ln |C| - \ln |C-B^{T} A^{-1} B|, \nn
&=& -71.49-(-72.97) = 1.48, 
\eea
where $\ln |A|$ contributes equally to competing AIC values and thus drops out. This removes the problem with machine precision leaving us a comparison of the logarithm of the determinant of smaller $25 \times 25$ matrices. We are now left with an easy calculation. The AIC changes by $\Delta \textrm{AIC} = \Delta  \ln |C_{\textrm{stat+sys}}|+\Delta \chi^2_{\textrm{min}} + 2 \Delta d = 1.5-1-6.2 +2 (5-2) = 0.3$, when one replaces the vanilla $\Lambda$CDM model with a (contradictory) $\Lambda$CDM model with a jump in the parameters $(H_0, \Omega_m)$ at $ z_{\textrm{split}} = 1$. Thus, despite the evolution seen in $(H_0, \Omega_m)$, Pantheon+ SNe data still has a marginal preference for the vanilla $\Lambda$CDM model over a physically ``ad hoc model'' with 3 additional parameters. }

{Given that our model $B$ is not only a contradiction, but also has 3 additional parameters, it is not really a serious contender. That being said, the take-home message is clear. If the $\Lambda$CDM fitting parameters $(H_0, \Omega_m)$ change with effective redshift in a statistically significant way (see later analysis in section \ref{sec:PD} for confirmation), thereby failing our consistency check for a given split into low and high redshift subsamples, this opens the door for competing models. A physically motivated minimal extension of the $\Lambda$CDM model evidently may lead to a reversal in the conclusion that the $\Lambda$CDM model is preferred. }

\begin{table*}[htb]
\centering 
\begin{tabular}{c|c|c|c|c|c|c|c|c}
\rule{0pt}{3ex} $z_{\textrm{split}}$ & \multicolumn{2}{c|}{\# SN} & \multicolumn{2}{c|}{$H_0$ (km/s/Mpc)} & \multicolumn{2}{c}{$\Omega_{m}$} & \multicolumn{2}{|c}{$\Delta \chi^2$}\\
\hline 
\rule{0pt}{3ex} & $\leq z_{\textrm{split}}$ & $> z_{\textrm{split}}$ & $\leq z_{\textrm{split}}$ & $> z_{\textrm{split}}$ & $\leq z_{\textrm{split}}$ & $> z_{\textrm{split}}$ & $\leq z_{\textrm{split}}$ & $> z_{\textrm{split}}$ \\
\hline
\rule{0pt}{3ex} $0.1$ &  $664$ &  $960$ & $73.19$ & $73.41$ & $0.359$ & $0.334$ & $-0.4$ & $0$\\
\rule{0pt}{3ex} $0.2$ & $871$ & $753$ & $73.19$ & $73.27$  & $0.388$ & $0.341$ & $-0.7$ & $-0.1$\\
\rule{0pt}{3ex} $0.3$ & $1130$ & $494$ & $73.24$ & $72.09$ & $0.374$ &  $0.384$ & $ -1.3$ & $-2.3$\\
\rule{0pt}{3ex} $0.4$  & $1316$ &  $308$ & $73.37$ & $72.64$ & $0.337$ &  $0.365$ & $0$ & $-0.3$  \\
\rule{0pt}{3ex} $0.5$  & $1414$ & $210$ & $73.38$ & $76.84$ & $0.333$ &  $0.252$ & $-0.1$ &  $-3.0$  \\
\rule{0pt}{3ex} $0.6$ & $1495$ &  $129$ & $73.30$ & $76.98$ & $0.348$ &  $0.249$ & $-0.5$ & $-1.0$  \\
\rule{0pt}{3ex} $0.7$ & $1549$ & $75$ & $73.30$ & {$80.29$} & $0.348$ &  {$0.190$} & $-0.5$ & $-2.4$  \\
\rule{0pt}{3ex} $0.8$ & $1594$ & $30$ & $73.27$ &  {$74.20$} & $0.353$ & {$0.266$} & $-1.1$ & $-1.7$  \\
\rule{0pt}{3ex} $0.9$ & $1597$ & $27$ & $73.26$ &  {$60.86$} & $0.354$ &  {$0.604$} & $-1.4$ & $-3.2$  \\
\rule{0pt}{3ex} {$1$} & {$1599$} & {$25$} & {$73.28$} &  {$34.37$} & {$0.351$} & {$3.391$} & {$-1.0$} &  {$-6.2$}  \\
\rule{0pt}{3ex} $1.1$ & $1604$ &  $20$ & $73.35$ &  {$34.19$} & $0.342$ &  {$3.478$} & $-0.3$ &  $-3.3$  \\
\rule{0pt}{3ex} $1.2$ & $1605$ &  $19$ & $73.37$ &  {$34.08$} & $0.340$ & {$3.508$} & $-0.1$ & $-2.5$  \\
\end{tabular}
\caption{{Redshift splits of the Pantheon+ sample showing the number of {SNe} (excluding 77 calibrators), the best fit $H_0$ and $\Omega_m$ values, and differences in $\chi^2$ in low and high redshift subsamples. Changes in $\chi^2$ are with respect to best fits for the full sample with no split (see Table \ref{tab:mock_input}). $M$ does not change as we decouple the calibrating {SNe} in the likelihood (\ref{chi2_split}). }}
\label{tab:zsplit_H0_Om}
\end{table*}

{We close this section with additional comments. The change of $(H_0, \Omega_m)$ parameters with effective redshift constitutes a decreasing $H_0$/ increasing $\Omega_m$ best fit trend with effective redshift. This is consistent with earlier analysis of the Pantheon SNe sample \cite{Colgain:2022nlb, Colgain:2022rxy}. Moreover, as is clear from Fig. 2 of \cite{Colgain:2022nlb} and Table \ref{tab:zsplit_H0_Om}, this trend begins at $z = 0.7$. Upgrading the Pantheon to Pantheon+ samples has not changed this trend. A final point worth stressing is that best fits beyond $z_{\textrm{split}} = 1$ prefer a $\Lambda$CDM model with negative DE densities, $\Omega_m > 1$. This is simply a feature of the Pantheon+ \cite{Brout:2022vxf, Scolnic:2021amr} data set, but since Risaliti-Lusso QSOs \cite{Risaliti:2018reu, Lusso:2020pdb} have a strong preference for $\Omega_m > 1$ inferences in the $\Lambda$CDM model at high redshifts, the observations are consistent and both data sets warrant further study.} 

\subsection{{An illustration of MCMC bias}}
{Having identified the split that enhances the improvement in fit, here we fix $z_{\textrm{split}}=1$ and present MCMC posteriors for data above and below the split. In Fig. \ref{fig:MCMCpriors} the results of this exercise can be seen, where we have allowed for different uniform priors on $\Omega_m$. There are a number of take-home messages. First, the low redshift ($H_0, \Omega_m$) posteriors are Gaussian, as expected, whereas the high redshift ($H_0, \Omega_m$) posteriors are not. Secondly, the peak of the $\Omega_m$ posterior is found in the $\Omega_m > 1$ regime, but it is robust to changes in the $\Omega_m$ prior. Thus, imposing $\Omega_m \leq 1$ would simply cut off the peak in the high redshift $\Omega_m$ posterior. Thirdly, the $H_0$ posterior is sensitive to the $\Omega_m$ prior. This is easy to understand as a projection effect. In short, as we relax the prior, the 2D MCMC posterior probes more of the top left corner of the $(H_0, \Omega_m)$-plane. Configurations in this corner only differ appreciably in $\Omega_m$, while getting projected onto more or less the same lower value of $H_0$. {Ultimately, what one concludes from the 2D MCMC posteriors is that the data is not good enough to constrain the model. The assumption then is that points in parameter space along the banana-shaped contour give rise to more or less the same values of $\chi^2$. As we shall show later, this assumption is false (see also \cite{Colgain:2023bge}). Of course, the peak of the marginalised 1D $H_0$ posterior} cannot be tracking the minimum of the $\chi^2$ as its value is unique up to machine precision. We will now introduce two independent methodologies, mock simulations and profile distributions, which track the minimum of the $\chi^2$, and we will assess the statistical significance of evolution between low and high redshift subsamples.}

\begin{figure}[htb]
\centering
\includegraphics[width=80mm]{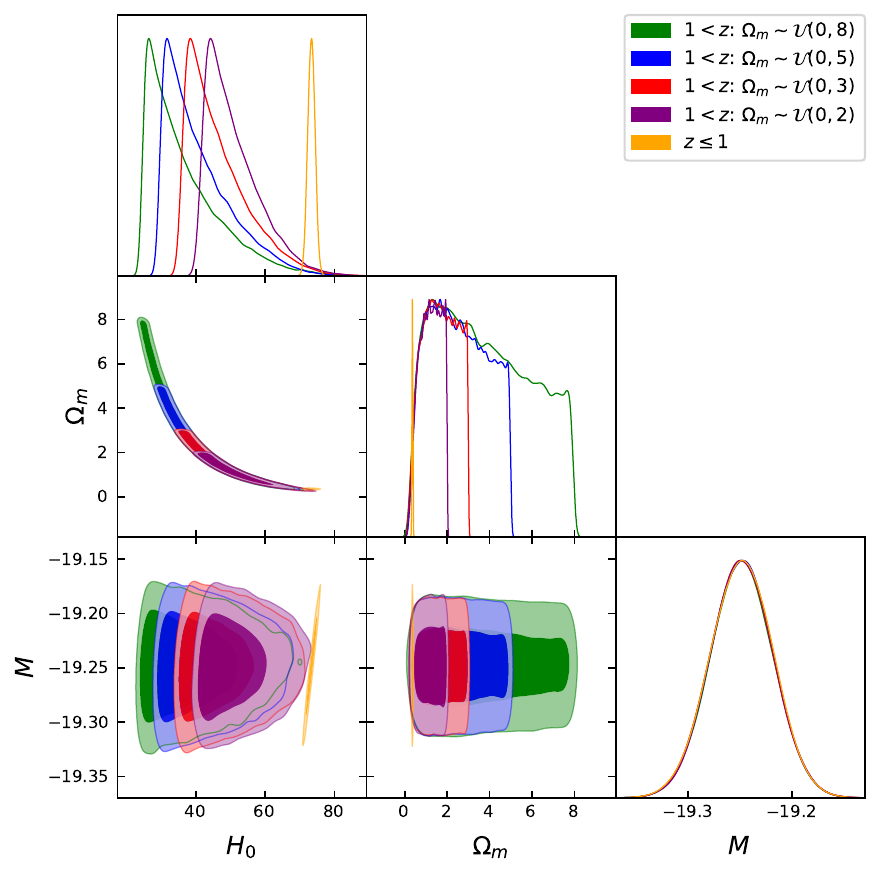} 
\caption{{MCMC posteriors for low and high redshift subsamples for the 2 cosmological parameters $(H_0, \Omega_m)$ and the 1 nuisance parameter $M$, the absolute magnitude of Type Ia SN. The low redshift posteriors are Gaussian, but the high redshift posteriors are not, in line with expectations. Extending the uniform $\Omega_m$ prior leads to shifts in the peak of the $H_0$ posterior due to a projection effect. Imposing the standard $\Omega_m \leq 1$ prior cuts off the peak of the $\Omega_m$ distribution in the high redshift subsample.}}
\label{fig:MCMCpriors} 
\end{figure}

\subsection{Frequentist Interpretation}
Here we adopt the same likelihood (\ref{chi2_split}), but estimate the probability of finding a decreasing $H_0$/increasing $\Omega_m$ best fit trend and negative DE densities as prominent in mock data. It should be noted that whenever one finds an unusual signal in cosmological data, it is standard practice to run mock simulations to ascertain if the signal is statistically significant or not. Here, the decreasing $H_0$/increasing $\Omega_m$ trend in best fits is the unusual signal that we wish to test. Since we search for evolution trends and one expects little evolution at low $z$ in mocks with good statistics, it is more efficient to remove low redshift {SNe} and restrict attention to the 210 {SNe} in the redshift range $z > 0.5$. Thus, given a realisation of {SNe} data, we choose a cut-off redshift $z_{\textrm{cut-off}}$ in the range $z_{\textrm{cut-off}} \in \{0.5, 0.6, 0.7, 0.8, 0.9, 1, 1.1, 1.2 \}$ and remove {SNe} with $z \leq z_{\textrm{cut-off}}$. This gives us 8 nested subsamples and for each subsample, we fit the $\Lambda$CDM model and record the best fit $(H_0, \Omega_m)$ values. We then construct the sums
\be
\label{sums}
{\sigma_{H_0} = \sum_{z_{\textrm{cut-off}}} ( H_0 - 73.41), \quad \sigma_{\Omega_m} = \sum_{z_{\textrm{cut-off}}} ( \Omega_m - 0.333),}
\ee
{where $H_0$ and $\Omega_m$ denote the best fits at each $z_{\textrm{cut-off}}$, and the difference is relative to the best fits of the full sample (Table \ref{tab:mock_input}). See \cite{Colgain:2022nlb} for earlier analysis with the Pantheon sample, where similar sums were employed but with a fixed (not fitted) $M$. Sums close to zero correspond to realisations of the data with no specific trend that averages to zero. As is clear from Table \ref{tab:zsplit_H0_Om}, in Pantheon+ we see a decreasing $H_0$ and increasing $\Omega_m$ trend, so we expect $\sigma_{H_0} < 0$ and $\sigma_{\Omega_m} > 0$ in Pantheon+ {SNe}; the concrete numbers are $\sigma_{H_0} = -115.50$ and $\sigma_{\Omega_m} = 9.27$ to two decimal places. The advantage of constructing a sum is that it places no particularly importance on the choice of $z_{\textrm{split}}$.}

\begin{table}[htb]
\centering 
\begin{tabular}{c|c|c}
 \rule{0pt}{3ex} $H_0$ (km/s/Mpc) & $\Omega_m$ & $M$ \\
\hline 
\rule{0pt}{3ex}  $ 73.41 \pm 1.04$ & $0.333 \pm 0.018$ & $-19.249 \pm 0.030$ \\
\end{tabular}
\caption{{Input parameters for our mocks. We construct an array of $(H_0, \Omega_m, M)$ values randomly in a normal distribution about each best fit with standard deviation specified by the error. Errors have been estimated through Fisher matrix.}}
\label{tab:mock_input}
\end{table}

Our goal now is to construct Pantheon+ {SNe} mocks in the redshift range $z > 0.5$ that are \textit{statistically consistent with no evolution in $(H_0, \Omega_m$)}. To begin, we fit the full sample using the likelihood (\ref{chi2_split}), identify best fits and $1 \sigma$ confidence intervals through the inverse of a Fisher matrix (see \cite{Perivolaropoulos:2023iqj}). We record the result in Table \ref{tab:mock_input}, noting that the result agrees almost exactly with (\ref{full_sample_constraints}), despite differences in the likelihood, i. e. (\ref{chi2}) versus (\ref{chi2_split}). {Note, we could also run an MCMC chain, but we have already demonstrated that the errors are Gaussian in section \ref{sec:warmup}, so whether one uses an MCMC chain or random numbers generated in normal distributions from Table \ref{tab:mock_input}, one does not expect a great difference.}

\begin{figure}[htb]
\centering
\begin{tabular}{c}
\includegraphics[width=90mm]{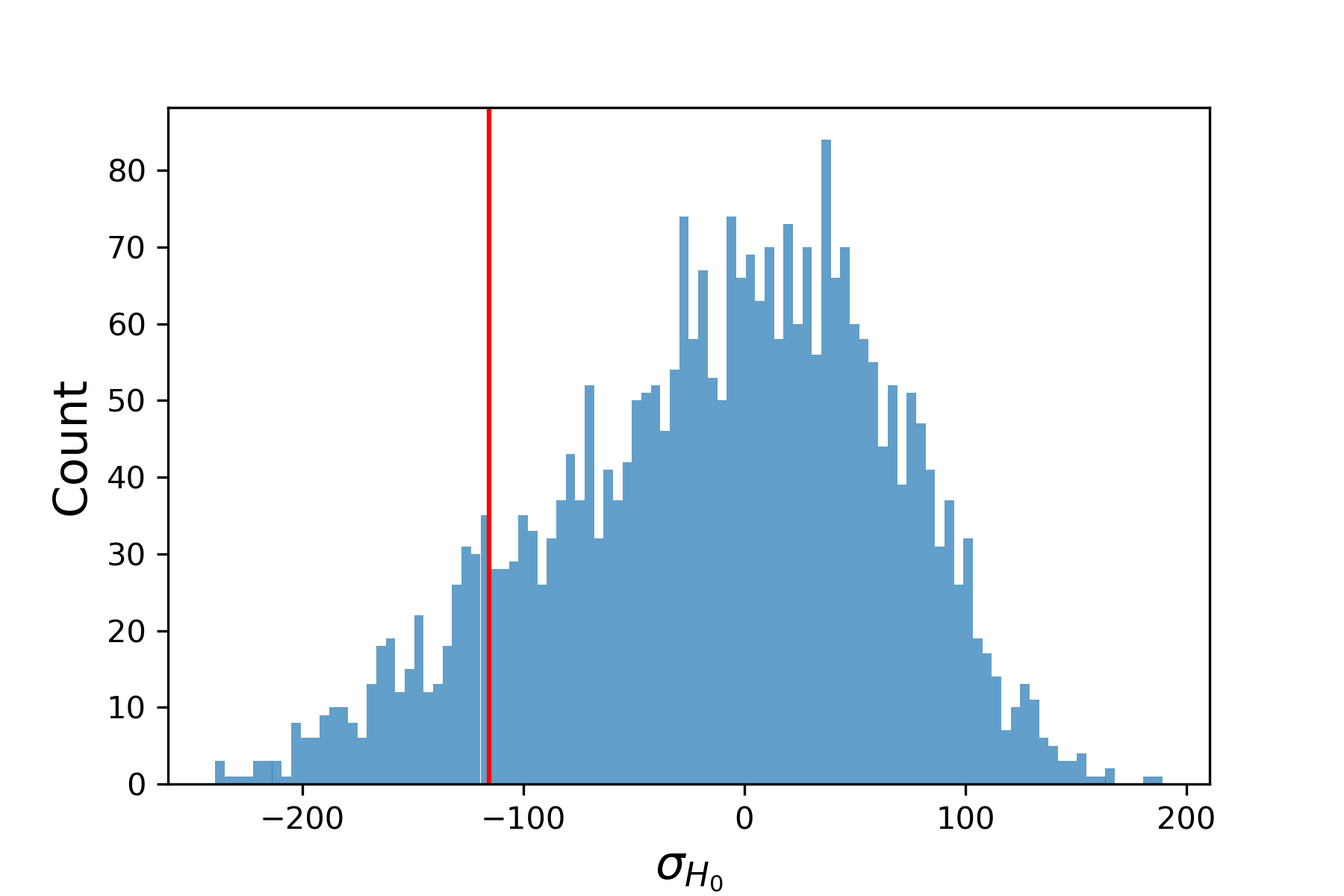} \\
\includegraphics[width=90mm]{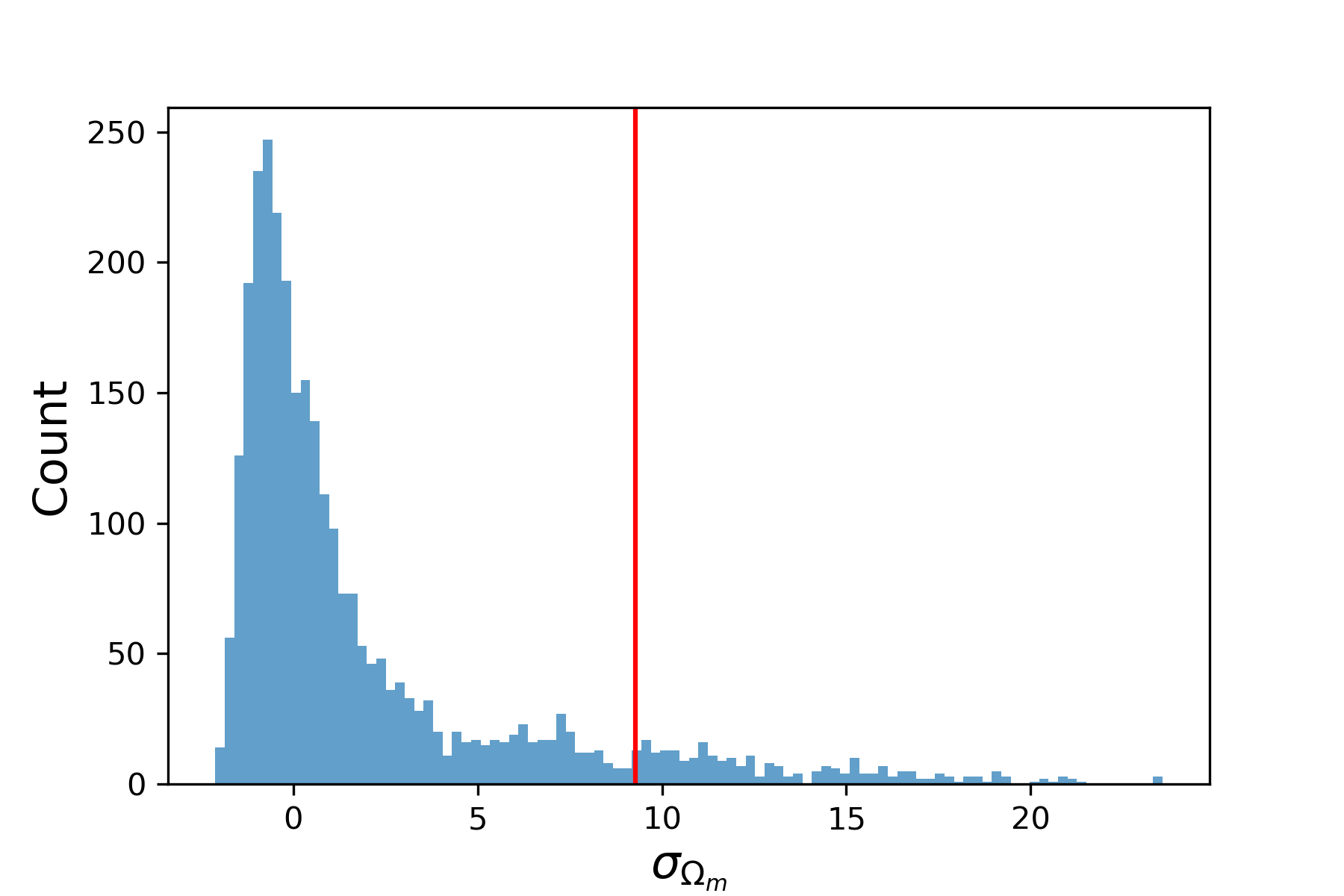}
\end{tabular}
\caption{{Sums (\ref{sums}) from 3,000 mocks where the input parameters were picked in normal distributions consistent with Table \ref{tab:mock_input}. Red lines denote the corresponding values from Pantheon+.} }
\label{fig:h0_om_sums_sim} 
\end{figure}

We next generate an array of 3,000 $(H_0, \Omega_m, M)$ values randomly in normal distributions with central value corresponding to the best fit value and $1 \sigma$ corresponding to the errors in Table \ref{tab:mock_input}. One could alternatively fix the injected cosmological paramaters to the best fits in Table \ref{tab:mock_input}, but our approach here allows for greater randomness. For each entry in this array, we construct $m_i = \mu_{\textrm{model}} (H_0, \Omega_m, z_i)+M$ for the 210 {SNe} in the redshift range $0.5 < z \leq 2.26137$. We then generate 210 new values of the apparent magnitude $m_i$ by generating a random multivariate normal with the covariance matrix $C_{\textrm{SN}}$ in (\ref{chi2_split_details}) truncated from the Pantheon+ covariance matrix $C_{\textrm{stat+sys}}$. This gives us one mock realisation of the data for each entry in our $(H_0, \Omega_m, M)$ array, which we fit back to the $\Lambda$CDM model for the nested subsamples in order to identify best fit parameters and the sums (\ref{sums}). Note, our mocking procedure drops correlations between $(H_0, \Omega_m, M)$, but this is not expected to make a big difference, since as can be seen from the yellow contour in Fig. \ref{fig:MCMCpriors}, which is representative of the full sample, none of the parameters are strongly correlated. Moreover, we do not generate new {SNe} in Cepheid hosts, so $M$ and its constraints are the same in mock and real data. This is justifiable because $M$ should be insensitive to cosmology \footnote{If this is not the case, then Type Ia {SNe} as \textit{standardisable} candles make little sense. Indeed, it should be safe to replace the 77 {SNe} in Cepheid hosts with a Gaussian prior on $M$.} and here our focus is studying evolution of $(H_0, \Omega_m)$ best fits in high redshift cosmological data. Once this is done for all 3,000 realisations, we count the number of mock realisations that give \textit{both} $\sigma_{H_0} \leq -115.50$ \textit{and} $\sigma_{\Omega_{m}} \geq 9.27$. Essentially, by ranking the mocks by $\sigma_{H_0}$ and $\sigma_{\Omega_m}$, one can assign a percentile or probability to the observed Pantheon+ sample, just as one would do with the heights of children in a class. Note, in both these exercises it is unimportant what the probability density function (PDF) looks like, simply that numbers are smaller or larger than a certain number. In Fig. \ref{fig:h0_om_sums_sim} we show the result of this exercise. As expected, our mock PDFs are peaked on $\sigma_{H_0} = \sigma_{\Omega_m} = 0$. From 3,000 mocks, we find 240 with more extreme values than the values we find in Pantheon+ SN. This gives us a $p$-value of $p = 0.08$ ($1.4 \sigma$ for a one-sided normal).

{We next consider the likelihood of finding negative DE densities ($\Omega_m > 1$), as well as the likelihood of finding $\Omega_m$ best fits as large as the Pantheon sample $ \Omega_m \gtrsim 3$, in the three final entries in Table \ref{tab:zsplit_H0_Om}. This can be done by recording best fit $\Omega_m$ values from mocks with $z_{\textrm{cut-off}} \in \{1.0, 1.1, 1.2\}$. From 3,000 mocks, we find  298 that maintain $\Omega_m > 1$ best fits and 77 that maintain larger $\Omega_m$ best fits than the Pantheon+ sample. This gives us probabilities of $p=0.1$ ($1.3 \sigma$) and $p=0.026$ ($1.9 \sigma$), respectively. In other words, we find negative DE densities in the same redshift range one mock in 10 and larger $\Omega_m$ best fits one mock in 38. In Fig. \ref{fig:omega_mocks} we show a subsample of the mock best fits.}

\begin{figure}[htb]
\centering
\includegraphics[width=90mm]{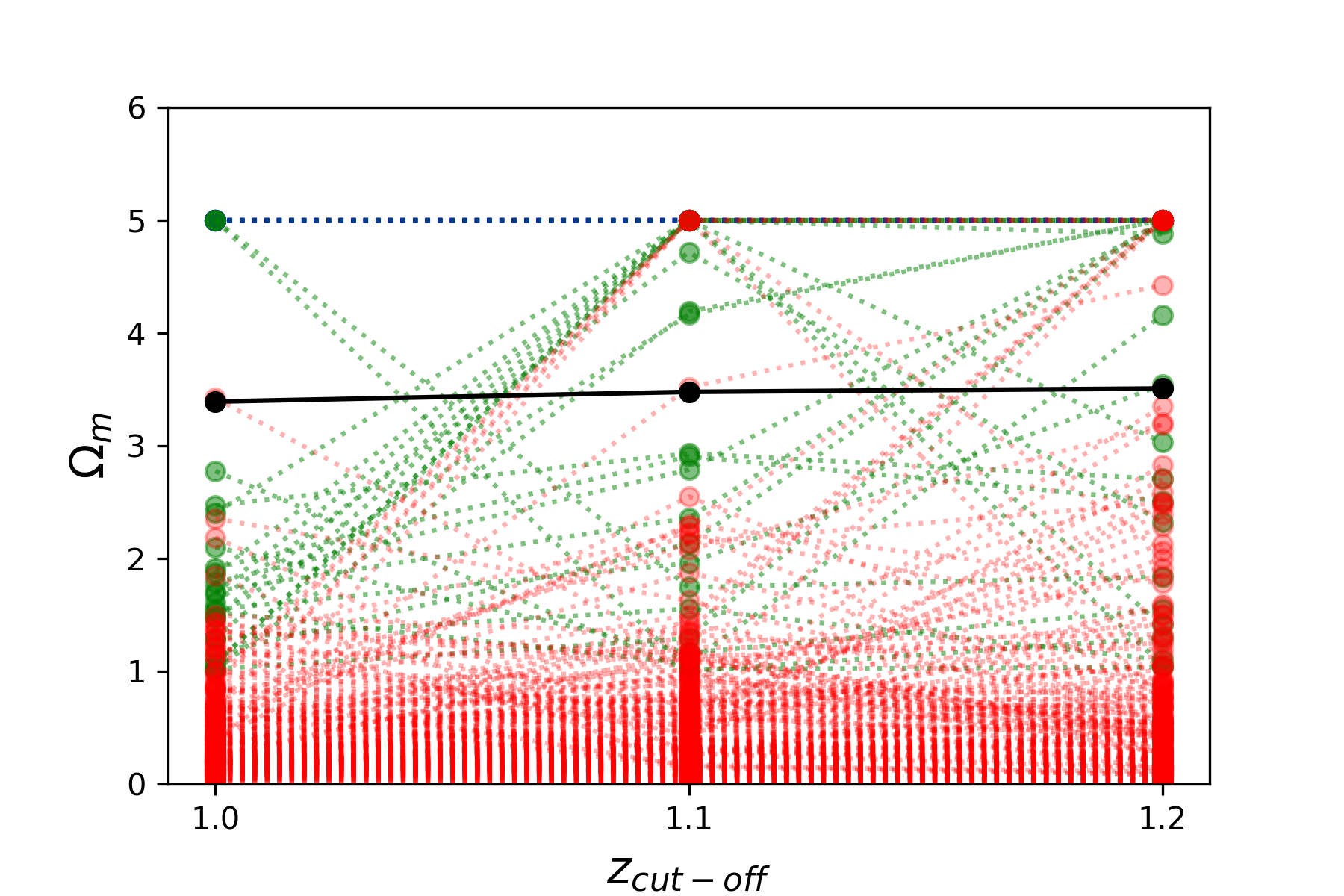} 
\caption{{A sample of 300 mock best fits for {SNe} with $z > z_{\textrm{cut-off}} \in \{1.0, 1.1, 1.2\}$. 6 mocks (blue) return $\Omega_m$ best fits that remain above the $\Omega_m$ best fits in Pantheon+ (solid black), 25 (green) that remain above $\Omega_m = 1$ and 269 (red) where best fits are recorded below $\Omega_m = 1$. We impose the bound $0 \leq \Omega_m \leq 5$ and some points saturate these bounds.}}
\label{fig:omega_mocks} 
\end{figure}
   
\subsection{{Pantheon+ Covariance Matrix \& $z_{\textrm{split}} = 1$}}
Here we comment on how representative are the high redshift best fit $(H_0, \Omega_m)$ values if one splits the sample at $z_{\textrm{split}}=1$. We perform this particular analysis so that we can directly compare to profile distributions in the next section. However, in the process we find a secondary result on the covariance matrix that is worth commenting upon. In contrast to the earlier sums, this means that we have singled out a particular redshift by hand and we are assessing the probability of a more specific event. For this reason we expect a probability less than $p = 0.08$. Once again we perform mock analysis, but surprisingly find that \textit{none of best fits to 10,000 mocks fits the data as well as the real data}. This may be partly due to the difference in preferred cosmological parameters, but is also expected to be due to a potential overestimation of the Pantheon+ covariance matrix \cite{Keeley:2022iba}.

\begin{figure}[htb]
\centering
\includegraphics[width=80mm]{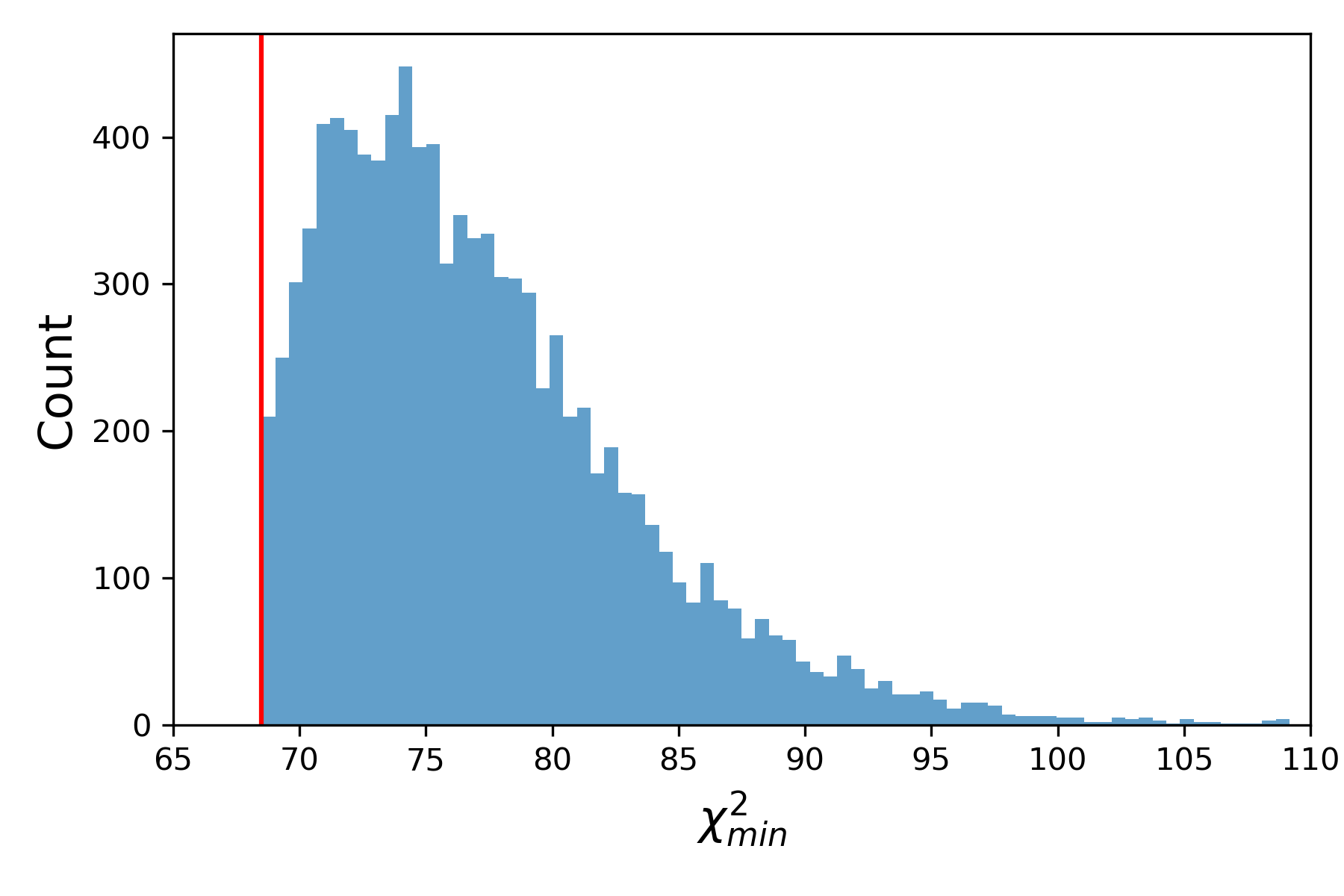} 
\caption{{Distribution of $\chi^2$ from 10,000 mocks of $z > 1$ {SNe} data, where input parameters are picked in normal distributions consistent with Table \ref{tab:mock_input}. The red line corresponds to the value in the Pantheon+ sample. None of our mock data result in smaller $\chi^2$ values than real data.}}
\label{fig:h0_om_sim} 
\end{figure}

Concretely, we construct an array of 10,000 $(H_0, \Omega_m, M)$ mock input parameters by employing the best fits and $1 \sigma$ confidence intervals in Table \ref{tab:mock_input} as central values and standard deviations for normal distributions. For each entry in this array, we generate a new mock copy of the 25 data points in the Pantheon+ sample above $z=1$, which we then fit back to the model and record 10,000 $(H_0, \Omega_m, M)$ best fits. Note, we are once again constructing high redshift subsamples that are representative of the full sample by construction. In Fig. \ref{fig:h0_om_sim} we show a comparison of the $\chi^2$ from mock data (blue PDF) versus $\chi^2$ from real data (red line); none of our mocks lead to lower values of $\chi^2$. Nevertheless, if one focuses on best fits, we find \textit{both} smaller values of $H_0$ \textit{and} larger values of $\Omega_m$ in 375 cases from 10,000 simulations, giving us a probability of $p = 0.0375$ of finding more extreme best fits. This corresponds to $ 1.8 \sigma$ for a one-sided normal. In the next section we will compare this statistical significance to profile distributions with the same data in the same redshift range. The key point here is that profile distributions provides a consistency check of our mock simulations. In short, if our mocks are trustworthy, we expect to see a $\sim 1.8 \sigma$ discrepancy in independent profile distribution analysis. A secondary point is that more extreme values of the $\chi^2$ are not found, which seems to support observations in \cite{Keeley:2022iba} that the Pantheon+ covariance matrix is overestimated.

\subsection{Restoring the Covariance Matrix}
Earlier we truncated out an off-diagonal block from the Pantheon+ covariance matrix in likelihood (\ref{chi2}) in order to decouple 77 {SNe} in Cepheid hosts from the remaining 1624 {SNe} and thus define the new likelihood (\ref{chi2_split}). Since this is heavy handed if one only wants to focus on high redshift SNe, here we restore the off-diagonal entries in the covariance matrix. The results are shown in Table \ref{tab:zsplit_cov}, where it is evident that the decreasing $H_0$/increasing $\Omega_m$ best fit trend with effective redshift is robust beyond $z_{\textrm{split}} = 0.7$. Moreover, we now find that {SNe} beyond $z_{\textrm{split}} = 0.9$ return best fits consistent with negative DE density. We have relaxed the bounds on $\Omega_m$ in order to accommodate best fits that saturate the bounds and the number of {SNe} excludes the 77 calibrating SNe. We also record a reduction in $\chi^2$ relative to the best fit values in Table \ref{tab:zsplit_H0_Om}, where the difference here is that we use (a truncation of) likelihood (\ref{chi2}) and not likelihood (\ref{chi2_split}). This provides a sanity check that our best fits are finding new minima as we change likelihood. Evidently, the re-introduction of off-diagonal entries in the covariance matrix impacts best fits, but not the features of interest. {As noted in the previous section, the Pantheon+ covariance matrix appears overestimated \cite{Keeley:2022iba}. }

\begin{table}[htb]
\centering 
\begin{tabular}{c|c|c|c|c}
 \rule{0pt}{3ex} $z_{\textrm{split}}$ & \# SN & $H_0$ (km/s/Mpc) & $\Omega_{m}$ & $\Delta \chi^2$ \\
\hline 
\rule{0pt}{3ex} $0.7$ & $75$ & $76.94$ & $0.250$ & $-0.6$ \\
\rule{0pt}{3ex} $0.8$ & $30$ & $69.73$ & $0.390$ & $-0.9$ \\
\rule{0pt}{3ex} $0.9$ & $27$ & $51.48$ & $1.148$ & $-0.9$ \\
\rule{0pt}{3ex} $1$ & $25$ & $20.28$ & $13.076$ & $-0.7$ \\
\rule{0pt}{3ex} $1.1$ & $20$ & $26.08$ & $7.248$ & $-0.5$ \\
\rule{0pt}{3ex} $1.2$ & $19$ & $28.64$ & $5.775$ & $-0.5$ 
\end{tabular}
\caption{{A repeat of the analysis in Table \ref{tab:zsplit_H0_Om} that includes additional off-diagonal terms in $C_{\textrm{stat+sys}}$ by only truncating the general likelihood (\ref{chi2}) to 77 {SNe} in Cepheid host galaxies and {SNe} with redshift $z > z_{\textrm{split}}$. $\Delta \chi^2$ is the difference between best fits in Table \ref{tab:zsplit_H0_Om}, where the likelihood (\ref{chi2_split}) was employed. We relaxed the bound $\Omega_m \leq 1$ to accommodate the best fit values.}}
\label{tab:zsplit_cov}
\end{table}

\section{Profile Distributions}
\label{sec:PD}
In this section we follow the methodology in \cite{Gomez-Valent:2022hkb}, more specifically \cite{Colgain:2023bge}, where we refer the reader for further details. As explained in \cite{Colgain:2022tql}, removing low redshift $H(z)$ or $D_{L}(z)$ or $D_{A}(z)$ data pushes the (flat) $\Lambda$CDM model into a non-Gaussian regime where projection effects are unavoidable. If one wants to test the constancy of $\Lambda$CDM cosmological parameters in the late Universe, and not simply resort to adopting a working assumption, then one has to overcome these effects. Profile distributions \cite{Gomez-Valent:2022hkb} allow one to construct probability density functions that are properly tracking the minimum of the $\chi^2$. The latter is by defintion the point in model parameter space that best fits the data. As is clear from Fig. \ref{fig:MCMCpriors}, where there is a degeneracy (banana-shaped contour) in the $(H_0, \Omega_m)$-plane, the peak of the $H_0$ posterior is sensitive to the prior, so it evidently tells one very little about the point in parameter space that best fits the data. {Note that profile distributions \cite{Gomez-Valent:2022hkb} are simply a variant of profile likelihoods (see section 4 of ref. \cite{Herold:2021ksg}), where instead of optimising one recycles the MCMC chain. As a result, the input for both Bayesian and frequentist analysis is the information in the MCMC chain, thereby allowing a more direct comparison between the two approaches.}

{Here we focus on $z_{\textrm{split}}=1$ as both our Bayesian and frequentist mock analysis suggests that this is the redshift split where evolution is most significant. Note, one can of course find sample splits with less evolution, but if one is interested in the self-consistency of a data set within the context of the $\Lambda$CDM model, it behoves us to focus on the most extreme cases. Following \cite{Gomez-Valent:2022hkb, Colgain:2023bge} we fix a generous uniform prior $\Omega_{m} \in [0, 8]$ and run a long MCMC chain for {SNe} with $z > z_{\textrm{split}} = 1$. The prior has been chosen large enough so that the expected best fit $\Omega_m \sim 3.4$ from Table \ref{tab:zsplit_H0_Om} ($z_{\textrm{split}} = 1$ row) can be recovered from the resulting distribution. We identify the minimum of the $\chi^2$, $\chi^2_{\textrm{min}}$, from the full MCMC chain. Next we break up the $H_0$ and $\Omega_m$ range into bins and record the lowest value of the $\chi^2$ in each bin, which gives us $\chi^2_{\textrm{min}}(H_0)$ and $\chi^2_{\textrm{min}}(\Omega_m)$, respectively. We can then define $\Delta \chi^2_{\textrm{min}}(H_0) := \chi^{2}_{\textrm{min}}(H_0) - \chi^2_{\textrm{min}}$ for $H_0$ and an analogous $\Delta \chi^2_{\textrm{min}}(\Omega_m)$ for $\Omega_m$. We next construct the distributions $R(H_0) = e^{-\frac{1}{2} \Delta \chi^2_{\textrm{min}}(H_0)}$ and $R(\Omega_m) = e^{-\frac{1}{2} \Delta \chi^2_{\textrm{min}}(\Omega_m)}$, which by construction are peaked at $R(H_0) = R(\Omega_m) = 1$ in the bin with the overall minimum of the $\chi^2$ for the full MCMC chain. }

\begin{figure}[htb]
\centering
\begin{tabular}{c}
\includegraphics[width=90mm]{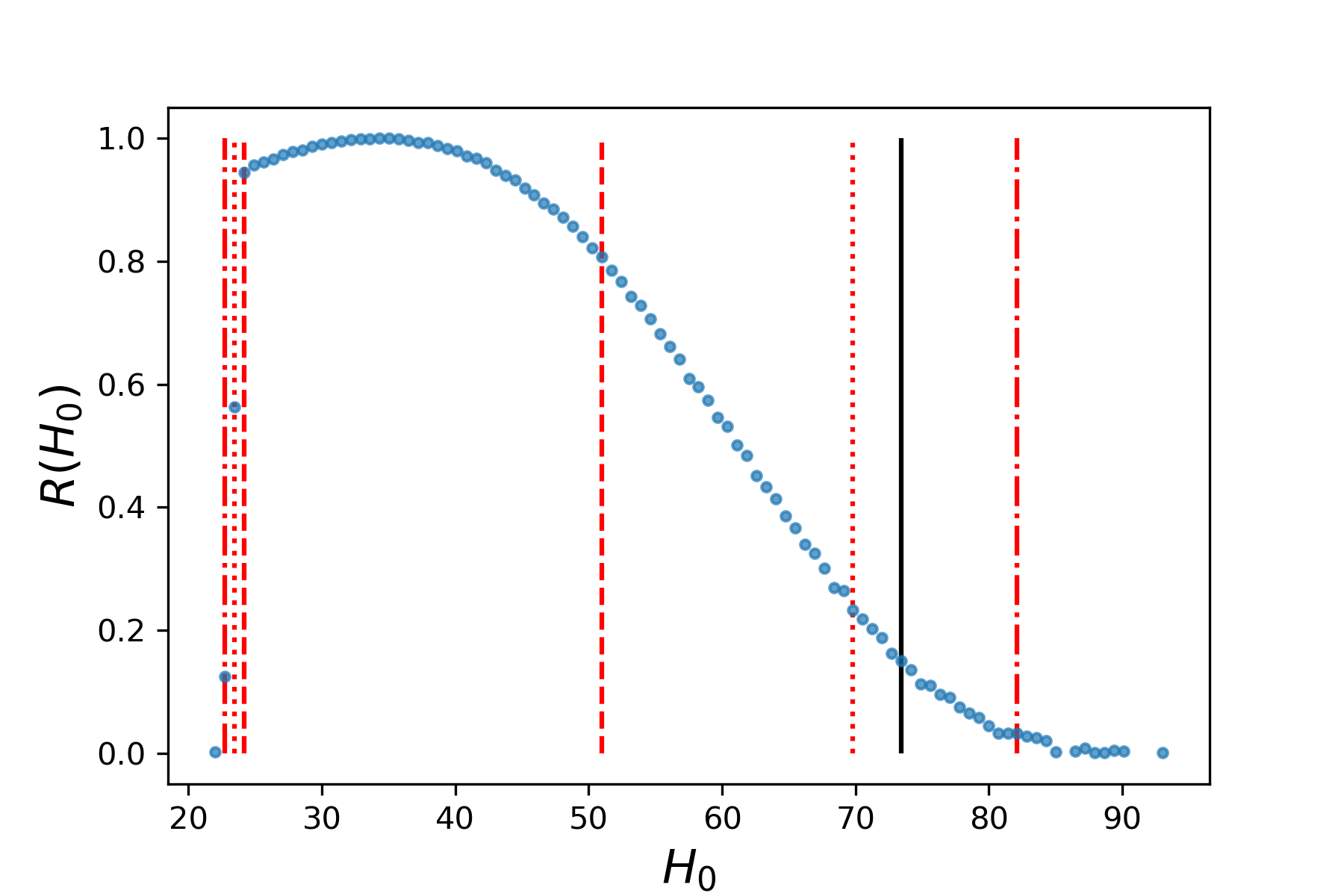} \\
\includegraphics[width=90mm]{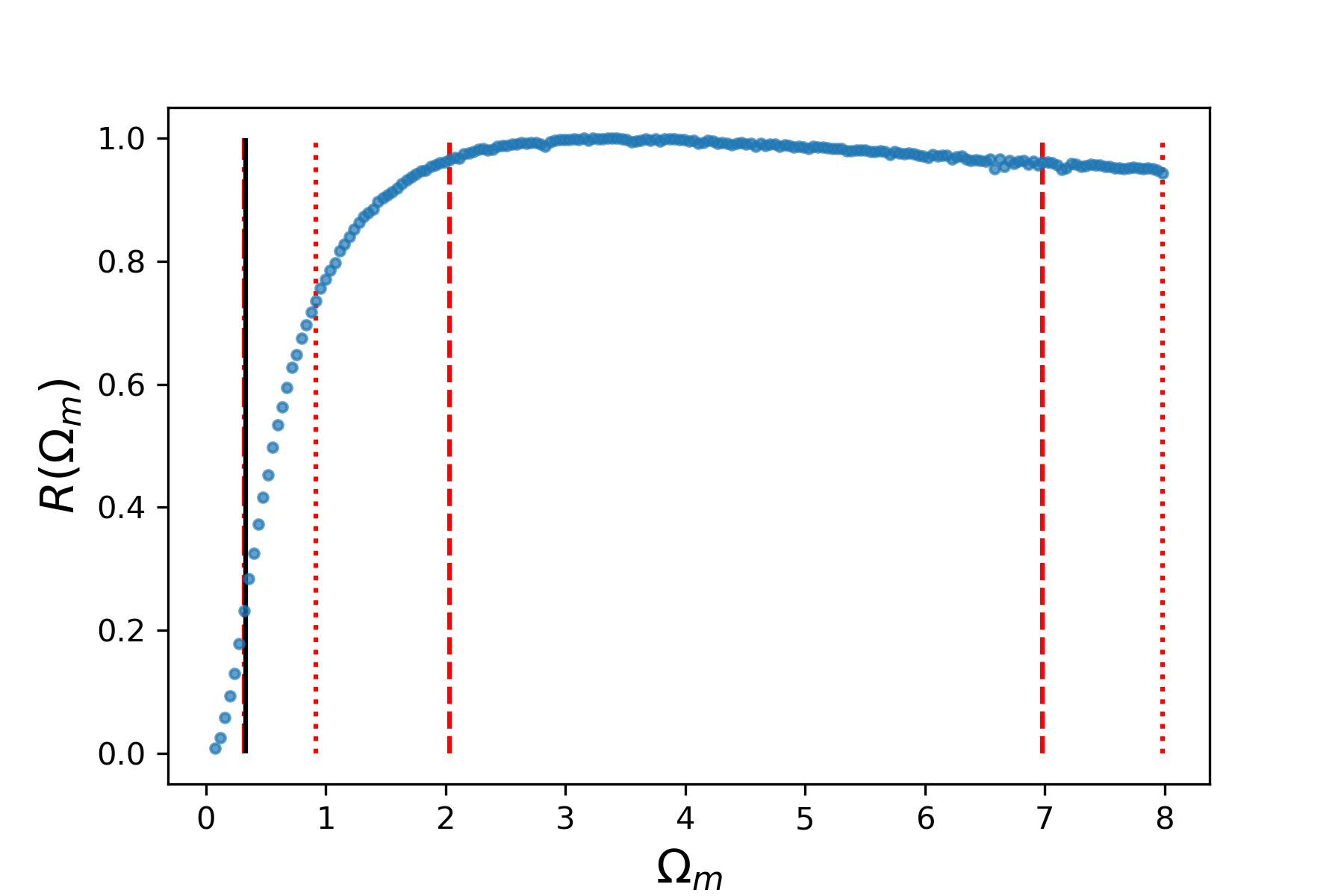}
\end{tabular}
\caption{{$R(H_0)$ and $R(\Omega_m)$ distributions for high redshift $z > 1$ {SNe} as a function of $H_0$ and $\Omega_m$. The black lines are the best fit values of the full Pantheon+ sample. Dashed, dotted and dashed-dotted lines denote $1 \sigma, 2 \sigma$ and $3 \sigma$, respectively.}}
\label{fig:PD} 
\end{figure}

It should be stressed that it is easy to select large enough priors for $H_0$ so that $R(H_0)$ decays to zero within the priors. Nevertheless, as is clear from Fig. \ref{fig:MCMCpriors}, $\Omega_m$ distributions become broad in high redshift bins and the fall off may be extremely gradual. However, once one switches from MCMC posteriors to profile distributions, we are no longer worried about the volume of parameter space explored in MCMC marginalisation, but simply that each bin is populated and the minimum of the $\chi^2$ in each bin has been identified. Thus, it is enough that the MCMC algorithm visits all bins at least once and any empty bin we omit. Concretely, we allow for 200 bins for both $H_0$ and $\Omega_m$. 

In Fig. \ref{fig:PD} we show the unnormalised $R(H_0)$ and $R(\Omega_m)$ distributions for high redshift {SNe} with $z > z_{\textrm{split}}=1$. The first point to appreciate is that the peaks of the distributions are close to the best fits in Table \ref{tab:zsplit_H0_Om}. Note, this provides a consistency check on the best fits, since extremising the $\chi^2$ through gradient descent and hopping around parameter space through MCMC marginalisation are independent. This provides a further test of the robustness of least squares fitting in this context (see also appendix). Secondly, as is evident from the dots to the left of the $R(H_0)$ peak, $R(H_0)$ goes to zero at both small and large values of $H_0$ that are well within our priors. In contrast, as anticipated, the $R(\Omega_m)$ distribution is almost constant beyond $\Omega_m \sim 2$ but nevertheless shows a gradual fall off. The fall off towards smaller values of $\Omega_m$ is considerably sharper. Thirdly, note that the dots essentially follow a curve, but some small bobbles are evident in bins. These features can be ironed out by running a longer MCMC chain. Finally, both $R(H_0)$ and $R(\Omega_m)$ confirm that the best fit for $z > 1$ {SNe} is not connected to the best fit for the full sample (black lines) through a curve of constant $\chi^2$. Thus, we see a degeneracy in (Bayesian) MCMC analysis, but there is no counterpart in a frequentist treatment that involves the $\chi^2$. We conclude that it is misconception in the literature that a degeneracy in MCMC posteriors is equivalent to a constant $\chi^2$ curve. We remind the reader again that the $\chi^2$ is a measure of how well a point in parameter space fits the data.

We next turn our attention to assessing the statistical significance. The black lines in Fig. \ref{fig:PD} denote the best fit values for the full sample from Table \ref{tab:mock_input}. Thus, these are the expected values if there is \textit{no evolution} in the sample. To assess the evolution, we normalise the $R(H_0)$ distribution by dividing through by the area under the full curve, which is most simply evaluated by numerically integrating using Simpson's rule. We then impose a threshold $\kappa \leq 1$ and retain only the $H_0$ bins with $R(H_0) > \kappa$. Integrating under the curve for the retained $H_0$ values and normalising accordingly one gets a probability $p$ \cite{Colgain:2023bge}. In Fig \ref{fig:PD} we use dashed, dotted and dashed-dotted lines to denote $p \in \{0.68, 0.95, 0.997 \}$ corresponding to $1 \sigma$, $2 \sigma$ and $3 \sigma$, respectively, in a Gaussian distribution. Evidently the best fit for the full sample (black line) is removed from the $H_0$ peak by a statistical significance in the $95 \%$ to $99.7\%$ confidence level range. By adjusting the threshold $\kappa$ further, one finds the area under the curve and the associated probability that terminates at the black line. 
%Converting this probability $p = 0.972$ into a more familiar Gaussian statistic, 
We find that the black line is located at the $97.2\%$ confidence level, the equivalent of $2.2 \sigma$ for a Gaussian distribution. This can be directly compared with $1.8 \sigma$ from our earlier analysis based on mock simulations. We note that there is a slight difference, but it is worth stressing that two independent techniques agree on a $\sim 2 \sigma$ discrepancy.

In principle one could repeat the analysis with $R(\Omega_m)$, but the distribution is broad and has been impacted by our priors. Changing the priors is expected to change the statistical significance of any inference using $R(\Omega_m)$, so we omit the analysis. If this is unclear, note that restricting the range to $\Omega_m \in [0, 4]$, would still allow a peak, but the dashed 
 and dotted lines corresponding to $68\%$ and $95\%$ of the area under the curve would all shift. The robust take-away is that the peak of the $R(\Omega_m)$ distribution coincides with negative DE density, $\Omega_m > 1$. However, there is an important distinction here with MCMC. As we see from Fig. \ref{fig:MCMCpriors}, due to a degeneracy in the 2D $(H_0, \Omega_m)$ posterior, changing the $\Omega_m$ priors can impact the $H_0$ posterior, whereas with profile distributions the number of times the MCMC algorithm visits a given $H_0$ bin is unimportant, simply the minimum $\chi^2$ in the $H_0$ bin is relevant. This important difference means that profile distributions are insensitive to changes in prior, modulo the fact that by changing the prior one either extends or cuts the distribution, but the peak does not move.

\section{Discussion} 
The take-home message is that a decreasing $H_0$/increasing $\Omega_m$ best fit trend observed in the Pantheon {SNe} sample \cite{Pan-STARRS1:2017jku} at low significance $\sim 1 \sigma$ \cite{Colgain:2022nlb} (see \cite{Krishnan:2020obg, Dainotti:2021pqg, Dainotti:2022bzg, Jia:2022ycc, Dainotti:2023yrk, Pasten:2023rpc} for the $H_0$ or $\Omega_m$ trend alone) persists in the Pantheon+ sample \cite{Brout:2022vxf, Scolnic:2021amr} with significance {$\sim 1.4 \sigma$ under similar assumptions that do not focus on a particular $z_{\textrm{split}}$}. Moreover, \textit{calibrated} $z >1$ {SNe} return $\Omega_m > 1$ best fits, thereby signaling negative DE densities in the $\Lambda$CDM model. Note, this outcome is not overly surprising, because one cannot preclude $\Omega_m > 1$ best fits at high redshifts even in mock Planck-$\Lambda$CDM data; beyond some redshift $\Omega_m > 1$ best fits become probable. This is a mathematical feature of the $\Lambda$CDM model \cite{Colgain:2022rxy, Colgain:2022tql}. Using profile distributions \cite{Gomez-Valent:2022hkb} (see also \cite{Colgain:2023bge}), a technique which allows us to correct for projection and/or volume effects in MCMC marginalisation, we have independently confirmed the significance at $\gtrsim 2 \sigma$. Similar features are evident in the literature, most notably Lyman-$\alpha$ BAO \cite{Aubourg:2014yra} and QSOs standardised through fluxes in UV and X-ray \cite{Risaliti:2015zla, Risaliti:2018reu, Lusso:2020pdb}. {Moreover, recent large SNe samples have led to larger $\Omega_m$ values that are $1.5 \sigma$ \cite{Rubin:2023ovl} to $2 \sigma$ \cite{DES:2024tys} discrepant with Planck \cite{Planck:2018vyg}. From Fig. 4 of ref. \cite{DES:2024tys} it is obvious that the sample has a high effective redshift. Note, in contrast to \cite{DES:2024tys}, where the high effective redshift is an inherent property of the sample, here we deliberately increase the effective redshift of the Pantheon+ sample by binning it.} 

To put these results in context we return to the generic solution of the Friedmann equation \cite{Krishnan:2020vaf},  
\be
\label{genericH}
H(z) = H_0 \exp \left(- \frac{3}{2} \int_0^{z} \frac{1+w_{\textrm{eff}}(z^{\prime})}{1+z^{\prime}} \textrm{d} z^{\prime} \right),  
\ee
where $w_{\textrm{eff}}(z)$ is the effective EoS. We observe that evolution of $H_0$ (and $\Omega_m$) with effective redshift in the Pantheon+ sample is consistent with a disagreement between the assumed EoS, here the $\Lambda$CDM model, and $H(z)$ inferred from Nature. These anomalies are not confined to {SNe} and we see related features elsewhere \cite{Wong:2019kwg, Millon:2019slk,Colgain:2022nlb, Colgain:2022rxy}. Moreover, JWST is also reporting anomalies that may be cosmological in origin \cite{Boylan-Kolchin:2022kae, Lovell:2022bhx, Menci:2022wia}; JWST anomalies may prefer a phantom DE EoS \cite{Menci:2022wia} {(however see \cite{Wang:2023ros,Adil:2023ara})}, which may be a proxy for negative DE densities at higher redshifts.
If persistent cosmological tensions \cite{Riess:2021jrx, Freedman:2021ahq, Pesce:2020xfe, Blakeslee:2021rqi, Kourkchi:2020iyz, DES:2021wwk, KiDS:2020suj} are due to systematics, one expects no evolution in $H_0$ from (\ref{genericH}), but this runs contrary to what we are seeing. {Our ``evolution test'', which may be regarded as a consistency check of the $\Lambda$CDM model confronted to data, hence gives a  complementary handle on establishing $\Lambda$CDM tensions, especially $H_0$ tension. Note, it is routine to fit data sets in cosmology and simply \textit{assume} that cosmological parameters are not evolving with effective redshift. Our analysis tests this \textit{assumption}.}

Admittedly, this one result may not be enough to falsify $\Lambda$CDM. {That being said, if evolution is present, as our Bayesian model comparison shows, this opens up the door for finding alternative models that fit the data better than vanilla $\Lambda$CDM. On the contrary, without any change of $(H_0, \Omega_m)$ with redshift across expansive Type Ia SNe samples, as is the standard assumption in the literature, there is little hope of finding an alternative that beats $\Lambda$CDM in Bayesian model comparison. From this perspective, our consistency check then feeds into standard Bayesian analysis. However, there is a key difference. Physics demands that models are predictive, i. e. return the same fitting parameters at all epochs, whereas Bayesian methods only assess the goodness of fit and are cruder. Note also that} the high redshift subsamples of Pantheon+ we study are small, so they are prone to statistical fluctuations. However, since we see similar trends beyond {SNe} \cite{Colgain:2022rxy}, this makes a statistical fluctuation interpretation less likely. A second possibility is unexplored systematics in $z > 1$ {SNe} identified largely through the Hubble Space Telescope (HST) \cite{SupernovaCosmologyProject:2011ycw, Riess:2017lxs, SupernovaSearchTeam:2004lze, Riess:2006fw}. There is unquestionable value in flagging these anomalies so that they can be explored. If one can eliminate these two, the only remaining possibility is that we must regard the trend as corroborating evidence that $\Lambda$CDM tensions are physical and the model is breaking down. 

Going forward, if the next generation of {SNe} data \cite{Scolnic:2019apa} increases the statistical significance of the anomaly, {as we have seen here in transitioning from Pantheon to Pantheon+}, then there are interesting implications. First, any increasing trend in $\Omega_m$ with effective redshift prevents one separating $H_0$ and $S_8 \propto \sqrt{\Omega_m}$ tensions. This is obvious. Interestingly, sign switching $\Lambda$ models, which perform well alleviating $H_0$/$S_8$ tensions \cite{Akarsu:2022typ}, fit well with our main message here, i.e. negative DE at higher redshifts. Secondly, $\Lambda$CDM model breakdown allows us to re-evaluate the longstanding observational cosmological constant problem \cite{Weinberg:1988cp}. Thirdly, and most consequentially, it is likely that changes to the DE sector cannot prevent evolution in $\Omega_m$, because DE is traditionally irrelevant at higher redshifts. Ultimately, if late-time DE does not or cannot come to the rescue \cite{Krishnan:2021dyb}, this brings the assumption of pressureless matter scaling as $a^{-3}$ with scale factor $a$ into question in late Universe FLRW cosmology. {Finally, if the evolution of $\Lambda$CDM parameters we discussed here is substantiated in future, it rules out the so-called early resolutions to $H_0$/$S_8$ tensions \cite{Abdalla:2022yfr}, such as early dark energy \cite{EDE}.}

\section{Acknowledgements}
We thank Leandros Perivolaropoulos for discussion, and  \"Ozgur Akarsu and Anjan Sen for comments on a late draft. MMShJ and SP  acknowledge SarAmadan grant No. ISEF/M/401332. SP also acknowledges the hospitality of Bu-Ali Sina University. 

%\begin{comment}
\appendix

\section{{Robustness of Least Squares Fitting}}
There is concern that least squares fitting may occasionally find local false minima. This issue can be addressed by starting the $\chi^2$ minimisation algorithm from different initial guesses (or priors) in parameter space. Here we show  differences that arise in such an exercise are insignificant. {Alternatively, as highlighted in the text, one could run an MCMC chain and identify the point in parameter space corresponding to the minimum of the $\chi^2$. If the resulting cosmological parameters agree well with $\chi^2$ minimisation, then this provides an additional check.}

\begin{table*}[htb]
\centering 
\begin{tabular}{c|c|c|c}
 \rule{0pt}{3ex} $(H_0, \Omega_m)$ & $ H_0$ (km/s/Mpc) & $\Omega_{m}$ & $\chi^2$ \\
\hline 
\rule{0pt}{3ex} $(\epsilon, \epsilon)$ & $73.284288$ & $0.35112504$ & $1491.11810412746$ \\
\rule{0pt}{3ex} $(\epsilon, 5-\epsilon)$ & $73.284282$ &  $0.35112510$ & $1491.11810412748$ \\
\rule{0pt}{3ex} $(150-\epsilon, \epsilon)$ & $73.284285$ & $0.35112503$ & $1491.11810412748$ \\
\rule{0pt}{3ex} $(150-\epsilon, 5-\epsilon)$ & $73.284290$ & $0.35112502$ & $1491.11810412746$ \\
\end{tabular}
\caption{{Parameters corresponding to the minima of the $\chi^2$ (\ref{chi2_split}) and minimum $\chi^2$ value for Cepheid host {SNe} and {SNe} in the range $0.00122 \leq z \leq 1$. The left column denotes the initial starting point with $\epsilon = 0.001$ for the algorithm at the four corners of our $(H_0, \Omega_{m})$ parameter space. As expected, differences in best fits are negligible.}}
\label{tab:testing_lsf_low}
\end{table*}

{We adopt uniform bounds or priors, $0 < H_0 < 150$ and $0 < \Omega_m < 5$, which are chosen large enough so that they never impact the best fits. As a result, the four corners of our parameter space are $(H_0, \Omega_m) = (\epsilon, \epsilon)$, $(H_0, \Omega_m) = (\epsilon,5-\epsilon)$, $(H_0, \Omega_m) = (150-\epsilon,\epsilon)$ and $(H_0, \Omega_m) = (150-\epsilon,5-\epsilon)$, where we adopt $\epsilon = 0.001$.  In Table \ref{tab:testing_lsf_low} we show the differences in best fit $H_0$, $\Omega_m$, and $\chi^2$ values for 77+1599 {SNe} in the low redshift range $0.00122 \leq z \leq 1$ when minimising the likelihood (\ref{chi2_split}). We truncate the resulting numbers where differences become transparent. As explained in the text, the parameter $M$ decouples, but throughout we start it from its best fit location $M = - 19.249$. From Table \ref{tab:testing_lsf_low} it is clear that the best fit $H_0$, $\Omega_m$ and the corresponding $\chi^2$ for the best fit, begin to differ at the  $5^{\textrm{th}}$, $7^{\textrm{th}}$ and $11^{\textrm{th}}$ decimal place, respectively.}

In Table \ref{tab:testing_lsf_high} we repeat the analysis for 77 {SNe} in Cepheid hosts and 25 {SNe} in the redshift range $1 < z \leq 2.26137$, where we see that differences begin at the $3^{\textrm{rd}}$, $4^{\textrm{th}}$ and $9^{\textrm{th}}$ decimal place, respectively. We have not changed the tolerance in the minimisation algorithm, but we see that the best fits in the smaller sample, where one expects less guidance for fits, are less robust. However, any difference is still small.

\begin{table}[htb]
\centering 
\begin{tabular}{c|c|c|c}
 \rule{0pt}{3ex} $(H_0, \Omega_m)$ & $ H_0$ (km/s/Mpc) & $\Omega_{m}$ & $\chi^2$ \\
\hline 
\rule{0pt}{3ex} $(\epsilon, \epsilon)$ & $34.366$ & $3.3914$ & $68.496048154$ \\
\rule{0pt}{3ex} $(\epsilon, 5-\epsilon)$ & $34.365$ & $3.3916$ & $68.496048156$ \\
\rule{0pt}{3ex} $(150-\epsilon, \epsilon)$ & $34.365$ & $3.3917$ & $68.496048156$ \\
\rule{0pt}{3ex} $(150-\epsilon, 5-\epsilon)$ & $34.365$ & $3.3916$ &  $68.496048156$ 
\end{tabular}
\caption{{Parameters corresponding to minima of the $\chi^2$ (\ref{chi2_split}) and minimum $\chi^2$ value for Cepheid host {SNe} and {SNe} in the range $1 < z \leq 2.26137$. The left column denotes the initial starting point with $\epsilon =0.001$ for the algorithm at the four corners of our $(H_0, \Omega_{m})$ parameter space. As expected, differences in best fits are negligible.}}
\label{tab:testing_lsf_high}
\end{table}

\end{document}